\documentclass[journal=nalefd,manuscript=letter,layout=twocolumn]{achemso}
\usepackage[version=3]{mhchem} 
\usepackage[utf8]{inputenc}
\usepackage[T1]{fontenc}
\usepackage{bm,bbold}
\usepackage{xcolor}
\usepackage{graphicx}
\usepackage{caption}
\usepackage{widetext}

\newcommand{\ket}[1]{\big|#1\big>}
\newcommand{\braket}[1]{\big<#1\big>}

\newcommand{\bk}{\mathbf{k}}

\newcommand{\bq}{\mathbf{q}}
\newcommand{\br}{\mathbf{r}}

\newcommand{\epstd}{\tilde\varepsilon}

\def\dz2{d$_{\text{z}^2}$}

\def\dx2y2{d$_{\text{x}^2\text{y}^2}$}
\def\G0W0{G$_0$W$_0$}
\def\scGW0{scGW$_0$}

\author{M.~Florian}
\affiliation
{Institut f\"ur Theoretische Physik, Universit\"at Bremen, P.O. Box 330 440, 28334 Bremen, Germany}
\email{mflorian@itp.uni-bremen.de}

\author{M.~Hartmann}
\affiliation
{Institut f\"ur Theoretische Physik, Universit\"at Bremen, P.O. Box 330 440, 28334 Bremen, Germany}

\author{A.~Steinhoff}
\affiliation
{Institut f\"ur Theoretische Physik, Universit\"at Bremen, P.O. Box 330 440, 28334 Bremen, Germany}

\author{J.~Klein}
\affiliation{Walter Schottky Institut and Physik Department, Technische Universit\"at M\"unchen, Am Coulombwall 4, 85748 Garching, Germany}

\author{A.~Holleitner}
\affiliation{Walter Schottky Institut and Physik Department, Technische Universit\"at M\"unchen, Am Coulombwall 4, 85748 Garching, Germany}
\altaffiliation{Nanosystems Initiative Munich (NIM), Schellingstr. 4, 80799 Munich, Germany}

\author{J.~J.~Finley}
\affiliation{Walter Schottky Institut and Physik Department, Technische Universit\"at M\"unchen, Am Coulombwall 4, 85748 Garching, Germany}
\altaffiliation{Nanosystems Initiative Munich (NIM), Schellingstr. 4, 80799 Munich, Germany}

\author{T.~O.~Wehling}
\affiliation
{Institut f\"ur Theoretische Physik, Universit\"at Bremen, P.O. Box 330 440, 28334 Bremen, Germany}
\altaffiliation{Bremen Center for Computational Materials Science, Universität Bremen, 28334 Bremen, Germany}

\author{M.~Kaniber}
\affiliation{Walter Schottky Institut and Physik Department, Technische Universit\"at M\"unchen, Am Coulombwall 4, 85748 Garching, Germany}
\altaffiliation{Nanosystems Initiative Munich (NIM), Schellingstr. 4, 80799 Munich, Germany}

\author{C.~Gies}
\affiliation
{Institut f\"ur Theoretische Physik, Universit\"at Bremen, P.O. Box 330 440, 28334 Bremen, Germany}

\title{The dielectric impact of layer distances on exciton and trion binding energies in van der Waals heterostructures}

\abbreviations{}

\begin{document}

%
%
%
%
%



\begin{abstract}
The electronic and optical properties of monolayer transition-metal dichalcogenides (TMDs) and van der Waals heterostructures are strongly subject to their dielectric environment. In each layer the field lines of the Coulomb interaction are screened by the adjacent material,
which reduces the single-particle band gap as well as exciton and trion binding energies.
By combining an electrostatic model for a dielectric hetero-multi-layered environment with semiconductor many-particle methods, we demonstrate that the electronic and optical properties are sensitive to the interlayer distances on the atomic scale. Spectroscopical measurements in combination with a direct solution of a three-particle Schr\"odinger equation reveal trion binding energies that correctly predict recently measured interlayer distances.


\end{abstract}


\paragraph{Introduction.}

\begin{figure*}[t!]
  \begin{center}
     \includegraphics[width=\textwidth]{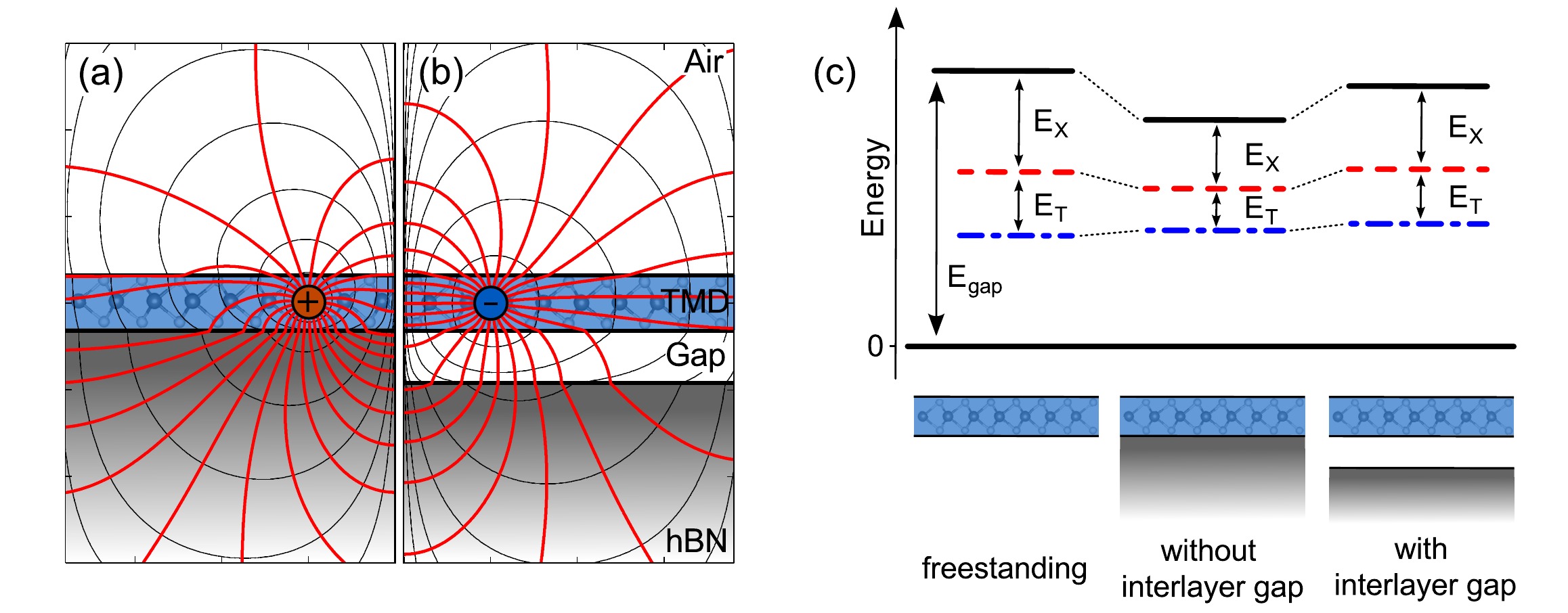}
    \caption{\textbf{(a)} Schematic representation of a vdWH with an ideal plane boundary and \textbf{(b)} a realistic interface with a finite interlayer gap. For two point charges in the TMD layer, calculated equipotential (black) and electric field lines (red) have been superimposed to visualize the effect of changes in the dielectric environment caused by the interlayer gap. \textbf{(c)} Illustration of the change of the band-gap and the bound-state energies due to the different screening environments: freestanding TMD monolayer (left), TMD monolayer on substrate with no distance between the layers causing strong screening  reducing both the band gap $E_{\text{gap}}$ and the binding energies of excitons $E_{\text{X}}$ and trions $E_{\text{T}}$ (middle), and non-vanishing gap between TMD and substrate leading to reduced screening (right).
     } \label{fig:scheme1}\label{fig:energies}
  \end{center}
\end{figure*}

Quantum-mechanical material design with atomically thin layers as the basic constituents is a relatively new discipline, driven by seemingly endless possibilities in material combinations in so-called van der Waals heterostructures (vdWH) \cite{geim_van_2013}, and by the manipulation and control of the electronic and optical properties due to their dielectric environment on the other hand. The Coulomb interaction between charge carriers in the atomically thin layer is screened only weakly, which is the reason for the exceptionally large exciton binding energies of hundreds of meV, and for the importance of GW corrections to band gaps calculated from density-functional theory \cite{huser_how_2013, qiu_optical_2013}. Field lines reaching out of the layer pass the surrounding dielectric environment (see Fig.~\ref{fig:scheme1}), and the effects of substrate screening have been heavily investigated in the recent years \cite{thygesen_calculating_2017, trolle_model_2017, raja_coulomb_2017, stier_probing_2016, ugeda_giant_2014, lin_dielectric_2014}. 
Moreover, the possibility of externally controlling the band gap and the binding energy is now recognized as a virtue to tailor new excitonic and optoelectronic devices by a dielectric encapsulation~\cite{gupta_direct_2017} and laser-annealing of metallic contacts~\cite{parzinger_contact_2017}. Prestructured substrates have been predicted \cite{rosner_two-dimensional_2016,steinke_non-invasive_2017} to induce lateral heterojunctions by local modification of the Coulomb interaction on the length scale of a few unit cells, which has recently been confirmed experimentally \cite{raja_coulomb_2017}. For vertically stacked heterostructures, often an intuitive picture is invoked that depicts field lines between two opposite charges in the 2D layer that pass the surrounding material. A realistic representation of the calculated electrostatic potential and the field lines is shown in Fig.~\ref{fig:scheme1}. While a quantitative assessment of the strength of the Coulomb interaction cannot be inferred from such a picture, it becomes clear that a difference in the dielectric environment, such as a finite gap at the heterostructure interface (right panel) instead of an ideal
plane boundary (left panel) does have an impact on the field lines. It is the topic of this letter to provide a quantitative understanding of the impact of the interlayer gap on the electronic and optical properties of vdWH.

VdWH consist of vertically stacked single layers of two-dimensional materials that can be semiconducting, such as MoS$_2$, MoSe$_2$, WS$_2$ and WSe$_2$, conducting, such as graphene, or insulating, such as boron nitride \cite{geim_van_2013,hong_interfacial_2017}. They can be fabricated under ambient conditions and without lattice matching due to the weak van der Waals interlayer bonding, which has led to an explosion of research activity on band-structure and interface engineering in this toolbox of materials. The atomistic modeling of heterostructures that are formed from incommensurate layers is strongly limited by computational demand. Different approaches have been developed to model the influence of the dielectric environment by invoking multiscale methods \cite{andersen_dielectric_2015,rosner_wannier_2015,latini_excitons_2015,trolle_model_2017,meckbach_influence_2017}.
These approaches have in common that an effective non-local dielectric function is obtained independently in a first step and is then successively used e.g.\ in Wannier-equation or BSE calculations to access the excitonic resonances for various substrates and substrate thicknesses \cite{trolle_model_2017,raja_coulomb_2017}. Alternatively, the optical response is obtained from the solution of semiconductor Bloch equations, which have been used before to evaluate the shift of  excitonic resonances in optically or electrically excited TMDs \cite{steinhoff_influence_2014,sie_observation_2017,berghauser_analytical_2014}.

Only recently, cross-sectional STEM \cite{rooney_observing_2017} and AFM \cite{tongay_tuning_2014} measurements have provided first insight into the actual layer separation at the interfaces, which is large enough (3 to 8\,\AA) to imply that Coulomb screening is significantly reduced by the gap between the layers. By introducing an electrostatic approach that builds on the \emph{Wannier function continuum electrostatics} (WFCE) scheme introduced in Ref.~\citenum{rosner_wannier_2015} to calculate the non-local dielectric function for an arbitrary number of stacked layers, we demonstrate a significant impact of realistic interface conditions on the non-local dielectric function that determines the screened Coulomb interaction -- and thereby the electronic and optical properties -- of vdWH.

To directly evaluate the impact of interlayer gaps at the interfaces in vdWH, we present a combined theoretical and experimental study of trion binding energies in various TMD/substrate combinations. Trion binding energies are calculated with sufficient accuracy to predict TMD-substrate layer separations in the experiment, which we find to be in agreement with recent cross-sectional STEM measurements \cite{rooney_observing_2017}. We further present results for the band-gap reduction and for the increase of exciton binding energies as function of the interlayer gap. A simple estimate is provided that allows for calculating corrections to the bound-state resonance energies for realistic interface conditions.
In the emerging field of band-structure and interface engineering in vdWH, our results demonstrate that layer separations at the interfaces strongly influence the long-range Coulomb interaction in the active layer and play an important role in the characteristics of optoelectronic devices.

\paragraph{Continuum electrostatics approach for calculating the non-local dielectric function in stacked layers.}

While the ease of fabrication is a particular benefit in creating vdWH, material-realistic ab initio calculations that are required to determine, amongst other things, band offsets and band gaps, quickly hit the computational limits, especially when supercells are required to represent incommensurable multi-layered materials. The result of recent efforts in the community has lead to different multi-scale approaches that share a common idea: While the electronic properties of the active TMD layer are determined from atomistic models, such as density functional theory \cite{qiu_optical_2013,huser_how_2013, huser_quasiparticle_2013} or effective tight-binding models \cite{rosner_two-dimensional_2016, steinke_non-invasive_2017,liu_three-band_2013,tong_topological_2016}, the dielectric screening that results from adjacent layers of various materials is treated in an electrostatic approach that is oblivious to the atomic resolution of each layer. This approach is based on the assumption that hybridization of
orbitals from adjacent layers plays a minor role as compared to dielectric screening. As long as we
concentrate on observables that emerge from the vicinity of the $K$ and $K$' points in reciprocal space that are well-protected against hybridization effects, this assumption is justified.

It is our aim to establish the importance of layer interfaces in vertical vdWH, in which the density of a polarizable medium is reduced due to the mere van der Waals interlayer bonding. As a consequence, field lines passing this ``interlayer gap'', as we will refer to it in the following, are more weakly screened in comparison to those passing the adjacent material. As we will show, the impact of the interlayer gap on the band gap and on excitonic binding energies is large due to the strong Coulomb effects in these materials. We use a two-step process to first provide closed equations for the non-local macroscopic two-dimensional
dielectric function $\varepsilon_\textrm{mac}^{\textrm{2D}}(\bq)$. The latter describes a TMD encapsulated in a sub- and superstrate heterostructure that includes additional layers of air to model the interlayer gaps. In a second step, this dielectric function is transformed into a microsopic basis and used to solve a generalized two- and three-particle Schr\"odinger equation to study the impact of interlayer distance on the exciton and trion binding energies.

In order to calculate properly screened Coulomb matrix elements for the embedded TMD, we begin with ab initio calculations for the freestanding monolayer to obtain bare $U_{\alpha\beta}(\mathbf q)$ and screened $V_{\alpha\beta}(\mathbf q)$ Coulomb matrix elements in a Wannier orbital basis $\ket{\alpha}$, where $\bq$ is a two-dimensional wave vector from the first Brillouin zone. To take into account environmental screening effects beyond the ab initio results, the central idea is that only the leading eigenvalue of $U$ is connected to long-wavelength charge-density modulations, for which environmental screening is expected to be strongest, while the remaining eigenvalues are linked to microscopic details and well described as constants. The same argument holds for the dielectric function, whose leading eigenvalue $\varepsilon_1(\mathbf q)=\varepsilon^{\textrm{2D}}_{\textrm{mac}}(\mathbf q)$ is hence analytically modelled by an effective two-dimensional dielectric function using continuum medium
electrostatics, as we show below. The matrix elements of the screened interaction $\mathbf{V}(\mathbf q)$ in
the eigenbasis of the bare interaction $\mathbf{U}(\mathbf q)$ are then obtained via $V_i(\mathbf q) = \varepsilon_i^{-1}(\mathbf q) \ U_i(\mathbf q)$ and transformed back into the Wannier basis.
To use them in equations of motion formulated in momentum space, they are subsequently transformed into the Bloch basis using expansion coefficients that connect the Wannier and the Bloch basis on a G$_0$W$_0$-level as described in Ref.~\citenum{steinhoff_influence_2014}.

The starting point of our derivation of a model dielectric function for TMD heterostructures is Poisson's equation, which yields the electrostatic potential $\phi(\br)$ for a given charge density $\rho(\br)$ in the presence of a dielectric function $\varepsilon_r(\br,\br')$ describing nonlocal screening effects \cite{trolle_model_2017}
\begin{align}
\nabla_{\br} \cdot\int d^3\br'\varepsilon_r(\br,\br') \nabla_{\br'}\phi(\br') = -\, \frac{\rho(\br)}{\varepsilon_0}.
\label{eq:poisson}
\end{align}
To find a unique solution for the potential $\phi$, we solve Poisson's equation for each layer of the heterostructure separately assuming an infinite extension of each layer in the x-y plane and a charge density $\rho$ only in the active TMD layer. At the interfaces, boundary conditions dictated by electrostatics must be fulfilled. To solve Eq.~(\ref{eq:poisson}), we transform the in-plane component to reciprocal space and use an ansatz for $\phi(\bq)$ that takes into account the vanishing of the potential at infinity and its continuity at each interface following from the continuity of the tangential electric field:
\begin{align}
\phi(\bq, z) = \frac{\rho(\bq)}{2\,\varepsilon_0\,\varepsilon_{\textrm{TMD}}(\bq)\,q}\, e^{-q\,|z|} \;+\; \sum_{j=1}^{N-1} B_j\, e^{-q\,|z-z_j|}\,.
\label{eq:ansatz}
\end{align}
Here, the first term accounts for the inhomogeneity due to the charge density in the active TMD layer. The second term stems from the homogeneous solutions of Poisson's equation in each layer, where $j$ runs over all interfaces of the $N$ layers. Its particular form captures the fact that surface charges given by the coefficients $B_j$  accumulate at each interface, thereby superimposing the two-dimensional Coulomb potential $\phi_0(\bq)/\varepsilon_{\textrm{TMD}}(\bq)=\rho(\bq)/(2\,\varepsilon_0\,\varepsilon_{\textrm{TMD}}(\bq)\,q)$ due to the charges in the active layer with induced potentials. Taking into account the combined action of the simple Coulomb potential and the induced potentials, we can formulate the two-dimensional dielectric function $\varepsilon^{\textrm{2D}}_\textrm{mac}(\bq)$ that describes the dielectric response to any charge density in the active TMD layer. The details of the electrostatic calculation are given in the Supporting Information and yield a system of coupled linear equations that can easily be solved for any relevant heterostructure size. In the following, we assume that the dielectric response of the active layer itself, $\varepsilon_{\textrm{TMD}}(\bq)$, is isotropic.

\begin{figure}[t]
\centering
\includegraphics[width=\columnwidth]{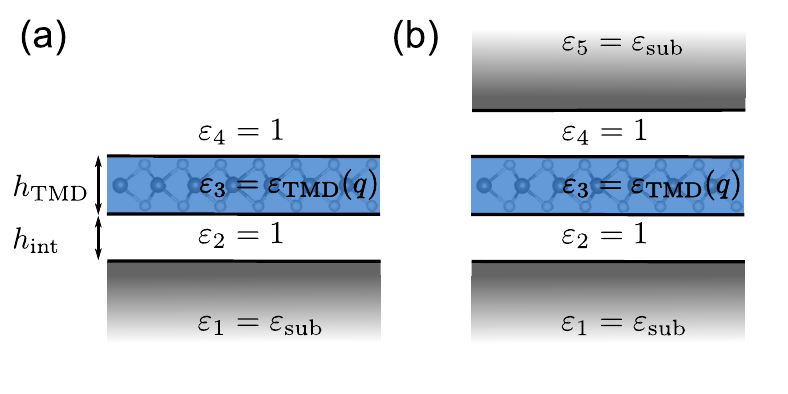}
\caption{Schematic respresentation of frequently encountered realizations of heterostructures accounting for an interlayer gap $h_\mathrm{int}$ between the active TMD layer and its surrounding. The effective non-local dielectric function for the supported and encapsulated cases \textbf{(a)} and \textbf{(b)} is given by Eqs.~\eqref{eq:eps_4schichten} and  \eqref{eq:eps_5schichten}, respectively.}
\label{fig:dielectric_situations}
\end{figure}

For the simple yet typical cases displayed in Fig.~\ref{fig:dielectric_situations} compact analytic expressions can be found. For a TMD layer of width $h_{\text{TMD}}$ placed on a substrate and accounting for the interlayer gap of width $h_{\textrm{int}}$ at the interface, we obtain

\begin{widetext}
\begin{align}
&\varepsilon^{\textrm{2D}}_\textrm{mac}(q) = \frac{\varepsilon_3(1 + \epstd_1\epstd_2 \beta + \epstd_1\epstd_3 \alpha^2\beta + \epstd_2\epstd_3 \alpha^2)}  {1 + \epstd_1 \alpha\beta + \epstd_2 \alpha - \epstd_3 \alpha + \epstd_1\epstd_2 \beta - \epstd_1\epstd_3 \alpha^2\beta - \epstd_2\epstd_3 \alpha^2 - \epstd_1\epstd_2\epstd_3 \alpha\beta}\,,
\label{eq:eps_4schichten}
\end{align}
\end{widetext}

\noindent
where we have defined $\epstd_i = \frac{\varepsilon_{i+1}-\varepsilon_i}{\varepsilon_{i+1}+\varepsilon_i}$ and $\alpha = e^{-q\, h_{\text{TMD}}}$, $\beta = e^{-q\, 2h_\text{int}}$ with the parameters $\varepsilon_1 = \varepsilon_\textrm{sub}$, $\varepsilon_2 = 1$, $\varepsilon_3 = \varepsilon_\textrm{TMD}(q)$, $\varepsilon_4 = 1$. Note that for a vanishing gap at the interface Eq.~\eqref{eq:eps_4schichten} reproduces the result of \cite{steinhoff_influence_2014, rosner_wannier_2015}. For the symmetric case of a TMD layer sandwiched between sub- and superstrate with equal dielectric constants $\varepsilon_\text{sub}$, the dielectric function is given by the simple expression
\begin{align}
\varepsilon_\textrm{mac}^{\textrm{2D}}(q) = \varepsilon_3 \, \frac{1-\epstd_1 \alpha\beta - \epstd_2 \alpha + \epstd_1\epstd_2 \beta}{1+\epstd_1 \alpha\beta + \epstd_2 \alpha + \epstd_1\epstd_2 \beta}~.
\label{eq:eps_5schichten}
\end{align}

\begin{figure}[t]
\centering
\includegraphics[width=\columnwidth]{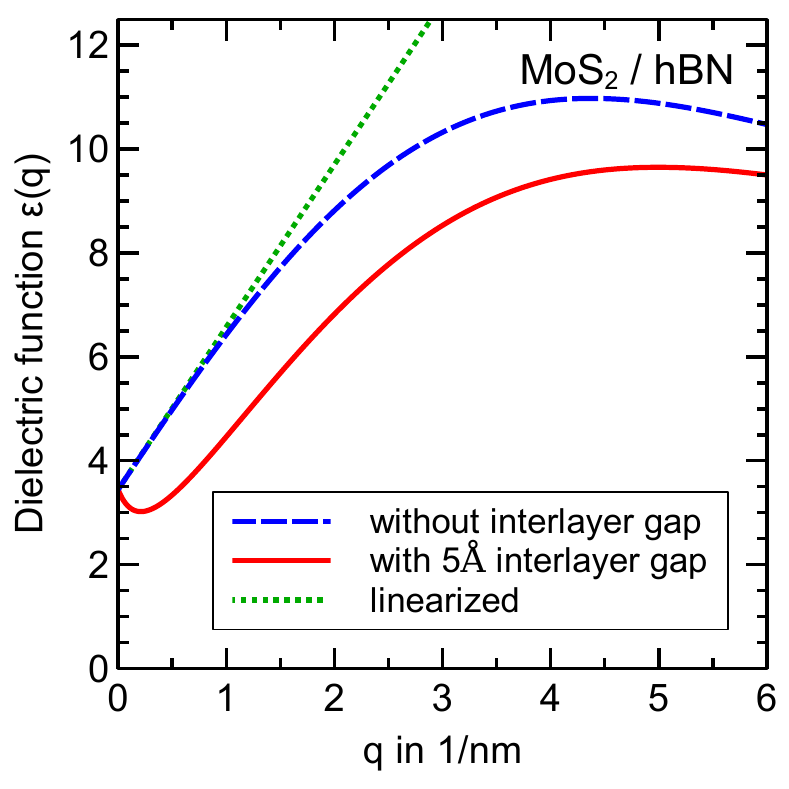}
\caption{Macroscopic dielectric function $\varepsilon_{\textrm{mac}}^{\textrm{2D}}(q)$ for MoS$_2$ on hBN, comparing results for an ideal plane (dashed line) interface and for a realistic interlayer gap of $h_\textrm{int}=5\,$\AA\,(solid line). The dotted line represents the linear behavior of the dielectric function if the Coulomb interaction is approximated by a Keldysh potential.}
\label{fig:eps_eff_example}
\end{figure}

In Fig.~\ref{fig:eps_eff_example} the impact of the substrate distance $h_\textrm{int}$ on the non-local dielectric function is shown for MoS$_2$ on top of hBN as obtained from Eq.~\eqref{eq:eps_4schichten}. Different regimes of screening can be identified depending on the wave vector $q$. In the long-wavelength limit ($q\rightarrow 0$), the effective screening is given by the average of substrate and superstrate dielectric constants in agreement with the Keldysh potential \cite{keldysh_coulomb_1979,berghauser_analytical_2014}. For small but finite momenta that are sensitive to the direct vicinity of the active TMD layer, the gap weakens the effective substrate screening and causes a pronounced dip below the long-wavelength value that is absent for $h_\textrm{int}=0$. For large momenta $q$, the effective dielectric function approaches the bulk limit, as it corresponds to charges being very close to each other in the TMD layer. The discrepancy between the cases with and without gap become particularly relevant in light of recent
results that have investigated interlayer gaps in vdWH that vary in different material classes and under annealing \cite{rooney_observing_2017,tongay_tuning_2014}. It further becomes clear that the material-realistic effective dielectric function is clearly beyond a linear description that is provided by a Keldysh potential (dotted line), and in using the latter one may strongly miscalculate the impact of screening.

\begin{figure}[t!]
\centering
\includegraphics[width=\columnwidth]{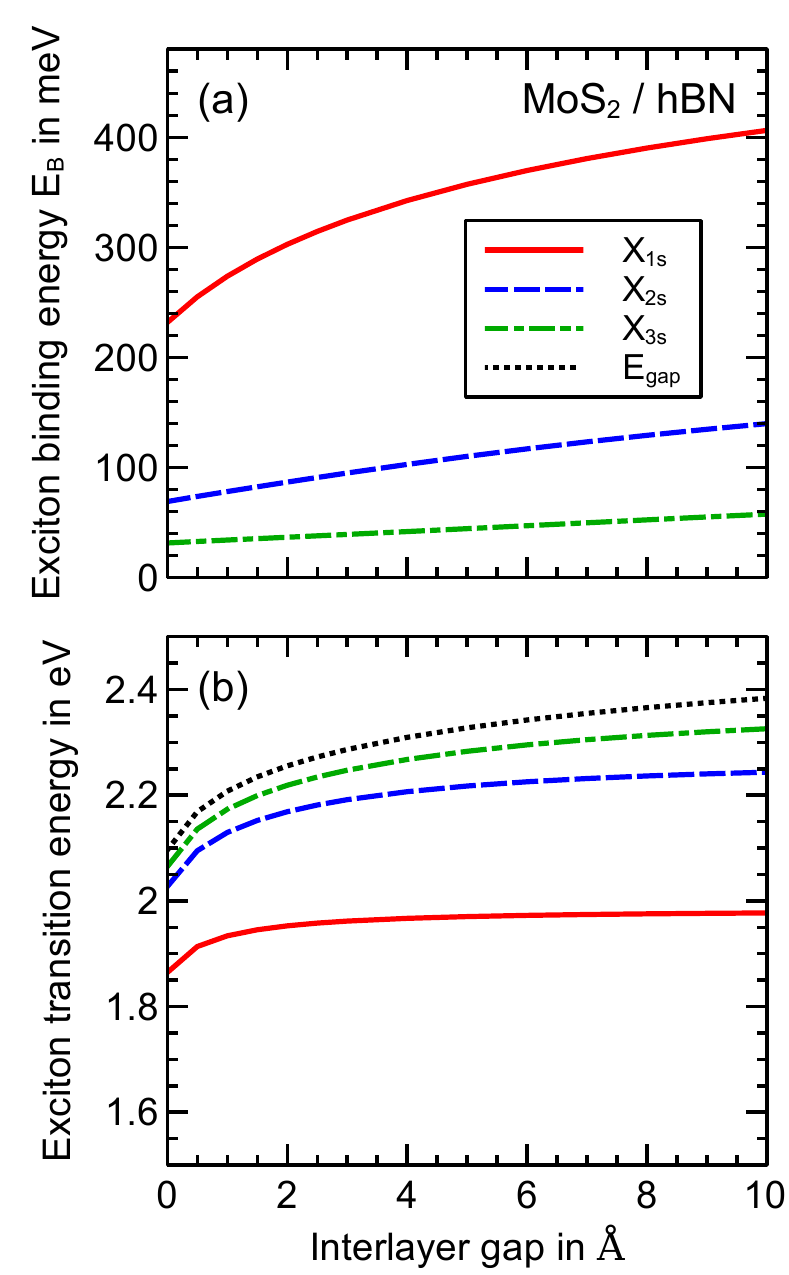}
\caption{\textbf{(a)} Impact of the interlayer gap on the binding energies of the exciton series in MoS$_2$ on an hBN substrate. \textbf{(b)} Absolute excitonic energies take into consideration the renormalized band gap, which is shown as a dotted line together with the energies of the 1s, 2s and 3s exciton transition. Further results for various combinations of TMDs and dielectric embeddings are given in the Supporting Information.}
\label{fig:binding_energies}
\end{figure}


\paragraph{Effect of interlayer distance on the two-particle optical properties and the band gap.}
The reduced screening in the presence of non-vanishing interlayer distances in vdWH significantly modifies the observable optical and single-particle properties of vdWH. Before we provide a direct assessment in terms of a theory-experiment comparison of the trion binding energy in the following section, we first take a look at the impact on the binding energies of the bound-state exciton series. Exciton states emerge as solutions of the semiconductor Bloch equations (SBE) for the microscopic interband polarisations $\psi^{he}_{\mathbf k} = \braket{a^h_{\mathbf k}a^{e}_{\mathbf k}}$ \cite{steinhoff_influence_2014,haug_quantum_1993}. In the limit of vanishing excitation density, the SBE become formally equivalent to the Bethe-Salpeter equation \cite{rohlfing_electron-hole_2000} and read in Fourier space
\begin{align}
    \begin{split}
        &(\varepsilon^e_{\mathbf k} + \varepsilon^h_{\mathbf k} - \hbar\omega)\psi^{he}_{\mathbf k}(\omega) \\
        & - \frac{1}{A}\sum_{\mathbf{k'}}\sum_{h'e'} V^{eh'he'}_{\mathbf k,\mathbf k',\mathbf k,\mathbf k'}\psi^{h'e'}_{\mathbf k'}(\omega) \\
        & = (d^{he}_{\mathbf k})^*E(\omega)\ .
    \end{split}
    \label{eq:SBE}
\end{align}
The linear response of the material is given by the macroscopic susceptibility $\chi(\omega) = \frac{1}{A}\sum_{\mathbf k}\sum_{he}\left( \mathrm d^{he}_{\mathbf{k}}\psi^{he}_{\mathbf k} + c.c.\right)/E(\omega)$, which contains excitons as discrete resonances below a continuum of optical interband transitions.
The screened Coulomb matrix elements $V$ and with them exciton binding energies $E_{\textrm{B}}$ depend directly on the dielectric function via the WFCE approach discussed above. Therefore, they are sensitive to modifications caused by variations in the interlayer gaps in vdWH discussed in the context of Fig.~\ref{fig:eps_eff_example}. We solve Eq.~\eqref{eq:SBE} by direct diagonalization using, in addition to properly screened Coulomb matrix elements, material-realistic input for band structures of the TMD slab as explained in detail in Ref.~\citenum{steinhoff_exciton_2017}.

In Fig.~\ref{fig:binding_energies}(a) the variation of the binding energy of the 1s  to 3s exciton resonances with interlayer gap is shown for the structure considered in Figs.~\ref{fig:scheme1}(b) and \ref{fig:eps_eff_example}. An increasing interlayer distance weakens the screening of the Coulomb interaction, which leads to a stronger electron-hole attraction and an increase of the binding energy.

A comparison of 1s to 3s exciton binding energies reveals that more tightly bound excitons are more susceptible to this effect. An understanding of this can be obtained by a series of approximations that is derived along the lines of Ref.~\citenum{olsen_simple_2016}. The central idea is to calculate an effective dielectric constant that is obtained by averaging $\varepsilon_{\textrm{mac}}^{\textrm{2D}}(q)$ over $|\bq|$ up to $1/a$, with $a$ being the exciton Bohr radius:
\begin{align}
    \varepsilon = 2a^2\int_0^{1/a}q\,\varepsilon_{\textrm{mac}}^{\textrm{2D}}(q)\;\mathrm d q\,.
    \label{eq:eff_dielectric_const}
\end{align}
Without the $q$-dependence of $\varepsilon$, the exciton problem can be solved analytically by means of the model of a 2D hydrogen atom. In this case the Bohr radius $a = \hbar^2\varepsilon/(2e^2\mu_{\text{ex}})$ is proportional to the dielectric constant and defines, together with Eq.~\eqref{eq:eff_dielectric_const}, a self-consistency problem.
Assuming that $\varepsilon_{\textrm{mac}}^{\textrm{2D}}(q)$ depends linearly on $|\bq|$ as in the case of a Keldysh potential, Eq.~\eqref{eq:eff_dielectric_const} can be solved analytically. The exciton binding energy is obtained by using the corresponding hydrogenic binding energy $E_{\text{B}}=\hbar^2/(2a^2\mu_{\text{ex}})$. In the presence of an interlayer gap the change of the binding energy can be obtained in the same spirit by means of first-order perturbation theory
\begin{align}
    \Delta E_{\text{B}}\approx \int_0^{1/a}q\Delta V(q)\;\mathrm d q\,,
\end{align}
where $\Delta V(q)=\frac{e^2}{4\pi\varepsilon_0 q}\Delta \varepsilon^{-1}(q)$ is the difference between the screened Coulomb potential with and without the interlayer gap. 

For the frequently used case of encapsulated TMDs with a dielectic function given by Eq.~\eqref{eq:eps_5schichten} an analytic expression can be derived in case of small interlayer gaps where $\Delta \varepsilon^{-1}(q) \approx \frac{\partial \varepsilon^{-1}(q)}{\partial h_{\text{int}}} h_{\text{int}}$. Assuming for simplicity that the dielectric response of the TMD layer itself $\varepsilon_\text{TMD}$ is momentum-independent and given by its bulk value, we obtain as a result:
\begin{align}
    \begin{split}
    &\Delta E_{\text{B}}\approx \frac{e^2}{4\pi\varepsilon_0}h_{\text{int}}\frac{4(\varepsilon_{\text{sub}}^2-1)}{h_{\text{TMD}}^2\varepsilon^+\varepsilon^-}\,\times\\
    &\times\left\{\left(1-\varepsilon^- \Lambda\right)\frac{h_{\text{TMD}}}{a} + \ln [(\varepsilon^++\varepsilon^-)\Lambda] \right\}\,,
    \end{split}
    \label{eq:Delta_EB_analytic_full}
\end{align}
with $\varepsilon^\pm = \varepsilon_{\text{sub}} \pm \varepsilon_{\text{TMD}}$ and $\Lambda=[\varepsilon^- + \varepsilon^+\exp{(h_{\text{TMD}}/a)}]^{-1}$. Taking advantage of the fact that the thickness of the TMD layer is small compared to the exciton Bohr radius ($h_{\text{TMD}}/a \ll 1$) Eq.~\eqref{eq:Delta_EB_analytic_full} reduces to a remarkably simple expression that is valid if the substrate screening is sufficiently strong ($\varepsilon_\text{sub}^2 \gg 1$):
\begin{align}
    \Delta E_{\text{B}}\approx \frac{e^2}{4\pi\varepsilon_0}\frac{h_{\text{int}}}{a^2}\,.
    \label{eq:Delta_EB_analytic}
\end{align}
For the asymmetric case of supported TMDs (cf.~Eq.~\eqref{eq:eps_5schichten}) we finally obtain the same result differing only by a factor of two which reflects the missing screening of the capping layer. From Eq.~\eqref{eq:Delta_EB_analytic} it becomes obvious that the exciton binding energy increases in the presence of an interlayer gap $h_{\text{int}}$. Eq.~\eqref{eq:Delta_EB_analytic} reveals further that excitons with larger Bohr radius $a$, such as 2s and 3s excitons, are less affected and the binding energy follows a characteristic $1/a^2$ dependence that we also obtain in the full calculation. It becomes clear that Coulomb effects are easily underestimated if material-realistic interlayer gaps between 3-8\AA\ are treated as ideal plane boundaries.


The absolute energy of the optical response of vdWH requires knowledge of the band gap in addition to the bound-state binding energies. In the SBE \eqref{eq:SBE} the exciton binding energy is affected by environmental screening via the Coulomb matrix elements $V^{eh'he'}_{\mathbf k,\mathbf k',\mathbf k,\mathbf k'}$, while the corresponding renormalization of the single-particle band gap has to be considered separately. The band structure of freestanding TMD slabs as obtained from $G_0 W_0$ calculations become modified since the long-range Coulomb interaction causing many-body renormalizations to the single-particle states experiences the very same environmental screening. This effect can be captured by a GdW self-energy, which was first brought up by Rohlfing\cite{rohlfing_electronic_2010} and used to describe screening-induced band structure renormalisation in vdWH\cite{winther_band_2017}. The idea is to approximately split the self-energy $\Sigma^{\textrm{GW,Het}}$ of the heterostructure into a part describing the
isolated TMD monolayer that is treated on a full ab-initio level, and a correction term containing environmental screening effects via a continuum-electrostatics model:
\begin{align}
\begin{split}
\Sigma^{\textrm{GW,Het}}& \approx G\,V^{\textrm{Het}} \\ &=G\,V^{\textrm{ML}}+G\,\Delta\,V=\Sigma^{\textrm{GW,ML}}+\Sigma^{\textrm{GdW}}
\label{eq:GdW}
\end{split}
\end{align}
with $\Delta\,V=V^{\textrm{Het}}-V^{\textrm{ML}}$. Here, the GdW self-energy leads to a correction of single-particle energies with respect to the monolayer band structure. To obtain the change of the band-gap energy in the presence of dielectric screening in vdWH, we evaluate the correction in static approximation, which leads to the screened-exchange-Coulomb-hole self-energy \cite{steinhoff_influence_2014} for conduction- ($c$) and valence-band ($v$) states
\begin{align}
\begin{split}
\Delta\,E_{\textrm{Gap},\bk}&=\Sigma^{\textrm{GdW},c}_{\bk}-\Sigma^{\textrm{GdW},v}_{\bk} \\ &=\sum_{\bk'}\Delta\,V_{\bk\bk'\bk\bk'}^{cccc}\left(\frac{1}{2}-f^{c}_{\bk'}\right)\\
&- \sum_{\bk'}\Delta\,V_{\bk\bk'\bk\bk'}^{vvvv}\left(\frac{1}{2}-f^{v}_{\bk'}\right) \\
&= \frac{1}{2}\sum_{\bk'}\left(\Delta\,V_{\bk\bk'\bk\bk'}^{cccc}+\Delta\,V_{\bk\bk'\bk\bk'}^{vvvv}\right)\,.
\label{eq:eps_GdW}
\end{split}
\end{align}
Here, we assume band-diagonal renormalizations with $f^{\lambda}_{\bk}$ being electron occupancies of the corresponding states. In the last step a full valence band ($f^{v} = 1$) and empty conduction band ($f^{c} = 0$) have been been considered. An evaluation of Eq.~\eqref{eq:eps_GdW} is easily performed, and the numerical results for band-gap energies as a function of the interlayer gap at the interface is shown by the dotted line in Fig.~\ref{fig:binding_energies}(b). In combination, the impact of the screened Coulomb interaction on the binding energies and the band gap leads to a significant shift of the bound-state optical transitions already for slight variations of the interlayer gaps in vdWH. To facilitate a direct comparison for various experimental realizations, the interlayer-gap dependence of the band gap and the exciton binding energies for further TMD/substrate combinations are provided in the Supporting Information.

\paragraph{Theory/Experiment Comparison of Trion Binding Energies.}

The energetic separation between the neutral and charged excitons (trions), here referred to as the trion binding energy $E_{\textrm{T}}$, is particularly well suited to study the effect of interlayer separation on Coulomb interaction. Signatures of tightly bound trion complexes are frequently observed in experimental spectra in the presence of moderate charge carrier densities~\cite{mak_tightly_2013,ross_electrical_2013,scheuschner_photoluminescence_2014}. As the difference between the trion and exciton bound-state energy, $E_{\textrm{T}}$ does not depend on the band gap and, therefore, directly reflects the strength of the Coulomb interaction and its screening, see Fig.~\ref{fig:energies}(c). Furthermore, it is more easily and with higher accuracy experimentally accessible in comparison to other methods that involve determining the separation between higher excited excitonic states~\cite{raja_coulomb_2017}, or combining optical measurements with single-particle measurements of the band gap~\cite{ugeda_giant_2014,ulstrup_ultrafast_2016}. In the following, we combine measurements of the trion binding energy with a solution of a generalized three-particle Schr\"odinger equation over the full
BZ. In combination with the electrostatic approach presented in the previous section, our model predicts trion binding energies with sufficient accuracy to extract layer separations in agreement with experimental results.


To access the trion, which is a three-particle property, the SBE \eqref{eq:SBE} are augmented by higher-order expectation values of the kind
\begin{align} \label{eq:trionampl}
    t^-_{e_1e_2h_3e_4}(\mathbf k_1, \mathbf k_2, \mathbf Q)=\braket{{a_{\mathbf{Q}}^{e_4}}^\dagger a_{-(\mathbf{k}_1+\mathbf{k}_2-\mathbf{Q})}^{h_3} a_{\mathbf{k}_2}^{e_2} a_{\mathbf{k}_1}^{e_1}}\,,
\end{align}
which are four-operator trion amplitudes. The particular one shown in Eq.~\eqref{eq:trionampl} is linked to the optical response of an electron trion $X^-$ that describes the correlated process of annihilating two electrons and one hole, leaving behind an electron with momentum $Q$ in the conduction band. Corresponding expressions for $t^+$ (hole trion $X^+$) can be obtained by utilizing the electron-hole symmetry. The trion amplitudes obey their own equation of motion, which we derive in first order in the carrier populations and in linear response~\cite{esser_theory_2001}: 

\begin{widetext}
\begin{align}
    \scriptstyle
        \begin{split}
        \scriptstyle
        &(\varepsilon^{e_1}_{\mathbf k_1} + \varepsilon^{e_2}_{\mathbf k_2} + \varepsilon^{h_3}_{\mathbf k_3}  - \varepsilon^{e_4}_{\mathbf Q} - \hbar\omega - i\Gamma)t^-_{e_1e_2h_3e_4}(\mathbf k_1, \mathbf k_2, \mathbf Q) \\
        &-\frac{1}{A}\sum_{\mathbf q}\sum_{h_5,e_6}V^{e_2h_5h_3e_6}_{\mathbf k_2,\mathbf k_3-\mathbf q,\mathbf k_3,\mathbf k_2-\mathbf q} t^-_{e_1e_6h_5e_4}(\mathbf k_1, \mathbf k_2- \mathbf q, \mathbf Q) \\
        &-\frac{1}{A}\sum_{\mathbf q}\sum_{h_5,e_6}V^{e_1h_5h_3e_6}_{\mathbf k_1,\mathbf k_3-\mathbf q,\mathbf k_3,\mathbf k_1-\mathbf q} t^-_{e_6e_2h_5e_4}(\mathbf k_1- \mathbf q , \mathbf k_2, \mathbf Q) \\
        &+\frac{1}{A}\sum_{\mathbf q}\sum_{e_5,e_6}V^{e_1e_2e_5e_6}_{\mathbf k_1,\mathbf k_2,\mathbf k_2+\mathbf q,\mathbf k_1-\mathbf q} t^-_{e_6e_5h_3e_4}(\mathbf k_1- \mathbf q , \mathbf k_2+\mathbf q, \mathbf Q) \\
&=f^{e_1}_{\mathbf{Q}}\left( \mathrm d^{he}_{\mathbf{k_2}} \delta_{\mathbf k_1,\mathbf Q} \delta_{e,e_1} - \mathrm d^{he}_{\mathbf{k_1}}\delta_{\mathbf k_2,\mathbf Q}  \right)E(\omega)\,.
    \end{split}
    \label{eq:eom_trion}
\end{align}
\end{widetext}
The homogeneous part of these equations is a generalization of a three-particle Schr\"odinger equation in reciprocal space for arbitrary band structures $\varepsilon^\lambda_{\bk}$ and Coulomb matrix elements $V^{\lambda_1\lambda_2\lambda_3\lambda_4}_{\bk_1,\bk_2\bk_3,\bk_4}(\bq)$.
The three-body problem determined by the SBE augmented by the trion amplitudes \eqref{eq:trionampl}, together with Eq.~\eqref{eq:eom_trion}, is solved by matrix inversion from which we obtain the linear absorption of the material by calculating the macroscopic susceptibility $\chi(\omega)$ as a response to the electric field propagating vertical to the heterostructure plane.

The optical response obtained from this approach contains both the bound-state trion and exciton resonances, and the trion binding energy is easily obtained from their energetic separation. We point out that our method is a material-realistic description on the full band structure and beyond both an effective mass approximation and a Keldysh potential for the Coulomb interaction that have been used in earlier works to calculate trion binding energies \cite{berkelbach_theory_2013, mayers_binding_2015, kylanpaa_binding_2015, velizhanin_excitonic_2015, ganchev_three-particle_2015, kidd_binding_2016, szyniszewski_binding_2017}. Especially the deviation from the linear behavior of the dielectric function displayed in Fig.~\ref{fig:eps_eff_example} clearly speaks against casting the Coulomb interaction into the shape of a simple Keldysh potential.

\begin{figure*}
  \begin{center}
    \includegraphics[width=.7\textwidth]{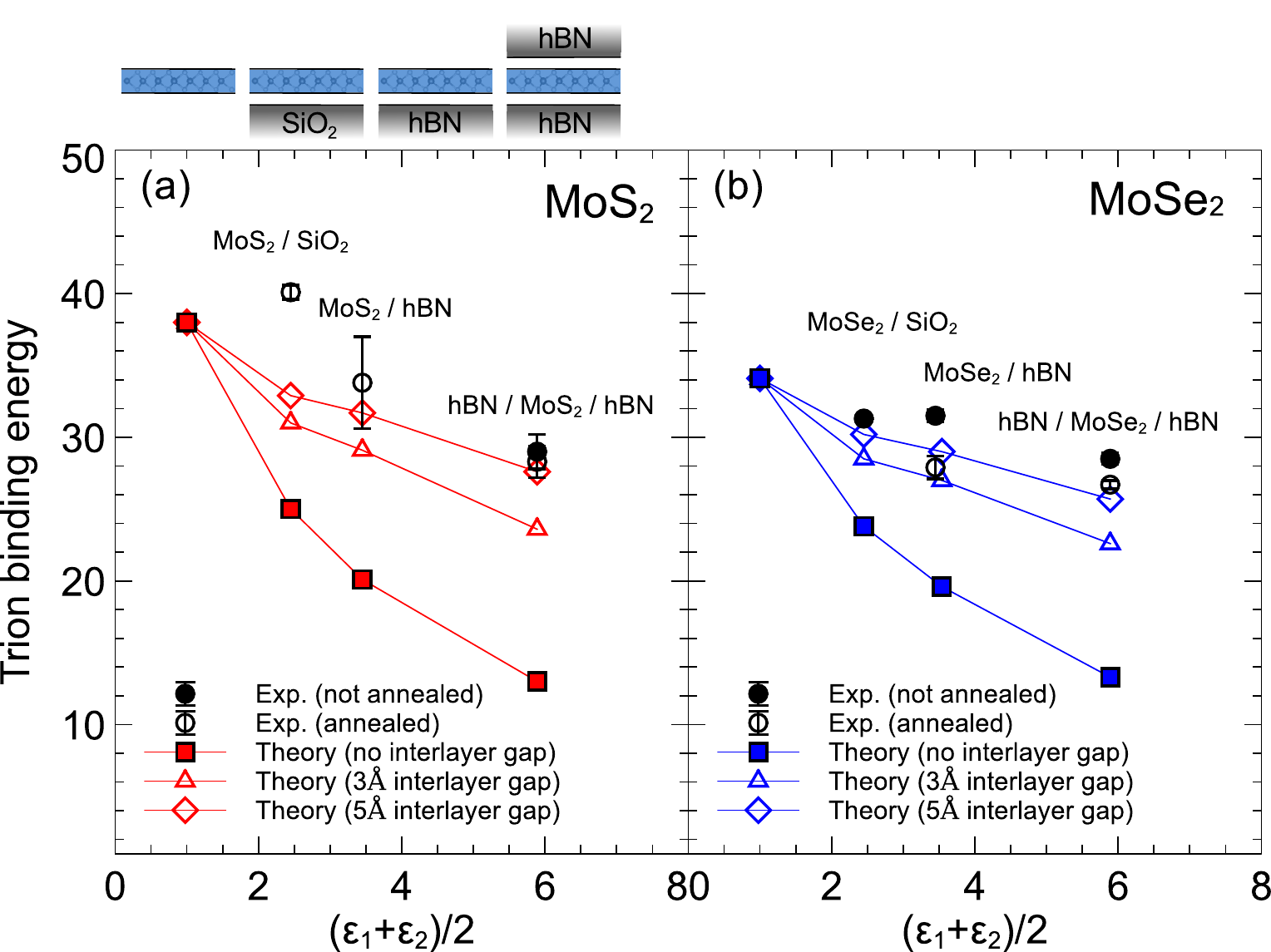}
    \caption{Trion binding energies determined from experiment for different vdWH before (closed symbols) and after (open symbols) annealing are shown together with theoretical results that have been obtained for corresponding structures and accounting for different sizes of interlayer gaps.} \label{fig:trionbindingenergies}
  \end{center}
\end{figure*}

To support our results on the sensitivity of Coulomb screening on the interlayer gap and to further demonstrate the accuracy of the trion binding energies obtained from our semiconductor model, we  present joint theory/experimental results for trion binding energies for various TMD/substrate combinations. The samples have been prepared by iteratively stacking hBN and TMD flakes by viscoelastic stamping onto a SiO$_{2}$/Si substrate. The thickness of hBN flakes used for stacking is typically of the order of 10-50\, nm. For the annealing step, samples are kept in a N$_{2}$-atmosphere at 50 mbar while being annealed at 300$^{\circ}$\, C for 30 minutes. In general, encapsulation with hBN and subsequent annealing results in almost lifetime-limited excitonic linewidths, with photostable photoluminescence for MoS$_{2}$ allowing to extract trion binding energies from low-temperature (10 K) photoluminescence spectra.~\cite{wierzbowski_direct_2017,ajayi_approaching_2017,cadiz_excitonic_2017}
By performing spatially resolved low-temperature (10 K) $\mu$-PL measurements and statistically analyzing emission spectra in different dielectric environments before and after annealing, we obtain trion binding energies. In all our measurements, we used continuous wave excitation at 2.33\, eV with an excitation power density of 0.3\,kWcm$^{-2}$.

Experimental results are shown as circles in Fig.~\ref{fig:trionbindingenergies} for intrinsically n-doped MoS$_{2}$ and MoSe$_{2}$ as a function of the long-wavelength limit of the dielectric screening induced by the dielectric embedding. Statistical errors are typically below 1\, meV reflecting high sample uniformity.
As expected for both TMDs a reduction of the trion binding energy is observed if the screening strength is successively increased by changing the substrate from SiO$_2$ to hBN and, further, by encapsulating the TMD in hBN. Annealing has been demonstrated before to be a crucial step in the fabrication of vdWH \cite{wang_interlayer_2016,tongay_tuning_2014}. By removing potential intercalated molecules, it has been shown that the interlayer separation is typically reduced by several \AA\ in the process of annealing. We observe a clear indications for a reduction of the binding energy of the electron trion after annealing (open circles), demonstrating that interlayer separation plays a noticable role for the Coulomb interaction strength. In fact, a microscopic calculation of the trion binding energy without an interlayer gap between the TMD and the sub-/superstrate (squares) strongly overestimates the experimentally observed binding energy reduction. Accounting for an interlayer gap within our electrostatic model of the dielectric screening using Eqs.~\eqref{eq:eps_4schichten} and \eqref{eq:eps_5schichten} we obtain quantitative agreement with the experimentally determined binding energies if a gap size of 3-5\,\AA\ is assumed (triangles, diamonds). This is in accordance with recent cross-sectional STEM measurements reporting an hBN-TMD interlayer distance of 5-7\,\AA\, which is measured between the atomically flat hBN layer and the TMD metal atom. A meaningful estimate for the interlayer gap is therefore obtained from the center-to-center interlayer distance by subtracting the metal-chalcogen vertical separation, which is of the order of 1.6 \AA~\cite{berkelbach_theory_2013}. Deviations are observed for MoS$_2$ on SiO$_2$, where the experimental binding energy is larger than the theoretical prediction. However, it has been argued~\cite{borghardt_engineering_2017} that water might be present on hydrophilic oxide surfaces. This leads to an additional layer of ice under cryogenic condition with a dielectric constant below 2 increasing the distance between TMD and substrate layer. The resulting binding energy might be compared with the theoretical result for a freestanding sample. This is supported by recent measurements on CVD grown MoS$_2$ on SiO$_2$ substrates, where trion binding energies of 35\,meV are reported \cite{cadiz_excitonic_2017}. Finally, relating our results to Monte-Carlo calculations \cite{mayers_binding_2015,kylanpaa_binding_2015} based on an effective-mass model for the band structure and a Keldysh potential shows that previously reported binding energies are underestimated by several meV (32.0-33.8 meV and 27.7-28.4 meV for free-standing MoS$_2$ and MoSe$_2$, respectively). Calculated binding energies of both the positively and the negatively charged trions in molybdenum- and tungsten-based TMDs and their dependence on the dielectric screening and the interlayer gap is provided in the Supporting Information.

\paragraph{Conclusion.}
We have investigated the impact of the distance between individual adjacent layers in vdWH with respect to Coulomb screening effects by combining a continuum electrostatic approach for calculating the non-local dielectric function in stacked vdWH with state-of-the-art semiconductor theory. A three-particle Schr\"odinger equation is solved on the basis of ab-initio input and on the full Brillouin zone to calculate trion binding energies. In a theory/experiment comparison we demonstrate the sensitivity of the Coulomb interaction to the interlayer distance and show that the accuracy of the calculations are sufficient to predict layer separations that are in excellent agreement with recent cross-sectional STEM measurements that reveal 5\AA\ layer-substrate separation for sulphur-based TMDs. Finally, we provide an approximate analytic expression for estimating the reduction of excitonic binding energies at the interfaces and  that reveals a $1/a^2$ scaling behavior with the Bohr radius $a$ of bound-state resonances. Our results may help explaining the variation in reported trion binding energies in the past~\cite{lee_identifying_2016} and  underlines the importance of accounting not only for layer thicknesses, but also for realistic conditions at the interfaces in the strongly evolving field of vdWH.

\begin{acknowledgement}
This work has been supported by the Deutsche Forschungsgemeinschaft via the graduate school ``Quantum-Mechanical Materials Modelling (QM$^3$)'' and through the ``TUM International Graduate School of Science and Engineering (IGSSE)''. We
gratefully acknowledge financial support of the
German Excellence Initiative via the ``Nanosystems
Initiative Munich'' and the PhD program ExQM of the Elite Network of Bavaria.
\end{acknowledgement}

\providecommand{\latin}[1]{#1}
\makeatletter
\providecommand{\doi}
  {\begingroup\let\do\@makeother\dospecials
  \catcode`\{=1 \catcode`\}=2\doi@aux}
\providecommand{\doi@aux}[1]{\endgroup\texttt{#1}}
\makeatother
\providecommand*\mcitethebibliography{\thebibliography}
\csname @ifundefined\endcsname{endmcitethebibliography}
  {\let\endmcitethebibliography\endthebibliography}{}


\begin{mcitethebibliography}{52}
\providecommand*\natexlab[1]{#1}
\providecommand*\mciteSetBstSublistMode[1]{}
\providecommand*\mciteSetBstMaxWidthForm[2]{}
\providecommand*\mciteBstWouldAddEndPuncttrue
  {\def\EndOfBibitem{\unskip.}}
\providecommand*\mciteBstWouldAddEndPunctfalse
  {\let\EndOfBibitem\relax}
\providecommand*\mciteSetBstMidEndSepPunct[3]{}
\providecommand*\mciteSetBstSublistLabelBeginEnd[3]{}
\providecommand*\EndOfBibitem{}
\mciteSetBstSublistMode{f}
\mciteSetBstMaxWidthForm{subitem}{(\alph{mcitesubitemcount})}
\mciteSetBstSublistLabelBeginEnd
  {\mcitemaxwidthsubitemform\space}
  {\relax}
  {\relax}

\bibitem[Geim and Grigorieva(2013)Geim, and Grigorieva]{geim_van_2013}
Geim,~A.~K.; Grigorieva,~I.~V. \emph{Nature} \textbf{2013}, \emph{499},
  419--425\relax
\mciteBstWouldAddEndPuncttrue
\mciteSetBstMidEndSepPunct{\mcitedefaultmidpunct}
{\mcitedefaultendpunct}{\mcitedefaultseppunct}\relax
\EndOfBibitem
\bibitem[Hüser \latin{et~al.}(2013)Hüser, Olsen, and
  Thygesen]{huser_how_2013}
Hüser,~F.; Olsen,~T.; Thygesen,~K.~S. \emph{Physical Review B} \textbf{2013},
  \emph{88}, 245309\relax
\mciteBstWouldAddEndPuncttrue
\mciteSetBstMidEndSepPunct{\mcitedefaultmidpunct}
{\mcitedefaultendpunct}{\mcitedefaultseppunct}\relax
\EndOfBibitem
\bibitem[Qiu \latin{et~al.}(2013)Qiu, da~Jornada, and Louie]{qiu_optical_2013}
Qiu,~D.~Y.; da~Jornada,~F.~H.; Louie,~S.~G. \emph{Physical Review Letters}
  \textbf{2013}, \emph{111}, 216805--216809\relax
\mciteBstWouldAddEndPuncttrue
\mciteSetBstMidEndSepPunct{\mcitedefaultmidpunct}
{\mcitedefaultendpunct}{\mcitedefaultseppunct}\relax
\EndOfBibitem
\bibitem[Thygesen(2017)]{thygesen_calculating_2017}
Thygesen,~K.~S. \emph{{2D} Materials} \textbf{2017}, \emph{4}, 022004\relax
\mciteBstWouldAddEndPuncttrue
\mciteSetBstMidEndSepPunct{\mcitedefaultmidpunct}
{\mcitedefaultendpunct}{\mcitedefaultseppunct}\relax
\EndOfBibitem
\bibitem[Trolle \latin{et~al.}(2017)Trolle, Pedersen, and
  Véniard]{trolle_model_2017}
Trolle,~M.~L.; Pedersen,~T.~G.; Véniard,~V. \emph{Scientific Reports}
  \textbf{2017}, \emph{7}, 39844\relax
\mciteBstWouldAddEndPuncttrue
\mciteSetBstMidEndSepPunct{\mcitedefaultmidpunct}
{\mcitedefaultendpunct}{\mcitedefaultseppunct}\relax
\EndOfBibitem
\bibitem[Raja \latin{et~al.}(2017)Raja, Chaves, Yu, Arefe, Hill, Rigosi,
  Berkelbach, Nagler, Schüller, Korn, Nuckolls, Hone, Brus, Heinz, Reichman,
  and Chernikov]{raja_coulomb_2017}
Raja,~A. \latin{et~al.}  \emph{Nature Communications} \textbf{2017}, \emph{8},
  15251\relax
\mciteBstWouldAddEndPuncttrue
\mciteSetBstMidEndSepPunct{\mcitedefaultmidpunct}
{\mcitedefaultendpunct}{\mcitedefaultseppunct}\relax
\EndOfBibitem
\bibitem[Stier \latin{et~al.}(2016)Stier, Wilson, Clark, Xu, and
  Crooker]{stier_probing_2016}
Stier,~A.~V.; Wilson,~N.~P.; Clark,~G.; Xu,~X.; Crooker,~S.~A. \emph{Nano
  Letters} \textbf{2016}, \emph{16}, 7054--7060\relax
\mciteBstWouldAddEndPuncttrue
\mciteSetBstMidEndSepPunct{\mcitedefaultmidpunct}
{\mcitedefaultendpunct}{\mcitedefaultseppunct}\relax
\EndOfBibitem
\bibitem[Ugeda \latin{et~al.}(2014)Ugeda, Bradley, Shi, da~Jornada, Zhang, Qiu,
  Ruan, Mo, Hussain, Shen, Wang, Louie, and Crommie]{ugeda_giant_2014}
Ugeda,~M.~M.; Bradley,~A.~J.; Shi,~S.-F.; da~Jornada,~F.~H.; Zhang,~Y.;
  Qiu,~D.~Y.; Ruan,~W.; Mo,~S.-K.; Hussain,~Z.; Shen,~Z.-X.; Wang,~F.;
  Louie,~S.~G.; Crommie,~M.~F. \emph{Nature Materials} \textbf{2014},
  \emph{13}, 1091--1095\relax
\mciteBstWouldAddEndPuncttrue
\mciteSetBstMidEndSepPunct{\mcitedefaultmidpunct}
{\mcitedefaultendpunct}{\mcitedefaultseppunct}\relax
\EndOfBibitem
\bibitem[Lin \latin{et~al.}(2014)Lin, Ling, Yu, Huang, Hsu, Lee, Kong,
  Dresselhaus, and Palacios]{lin_dielectric_2014}
Lin,~Y.; Ling,~X.; Yu,~L.; Huang,~S.; Hsu,~A.~L.; Lee,~Y.-H.; Kong,~J.;
  Dresselhaus,~M.~S.; Palacios,~T. \emph{Nano Letters} \textbf{2014},
  \emph{14}, 5569--5576\relax
\mciteBstWouldAddEndPuncttrue
\mciteSetBstMidEndSepPunct{\mcitedefaultmidpunct}
{\mcitedefaultendpunct}{\mcitedefaultseppunct}\relax
\EndOfBibitem
\bibitem[Gupta \latin{et~al.}(2017)Gupta, Kallatt, and
  Majumdar]{gupta_direct_2017}
Gupta,~G.; Kallatt,~S.; Majumdar,~K. \emph{Physical Review B} \textbf{2017},
  \emph{96}, 081403\relax
\mciteBstWouldAddEndPuncttrue
\mciteSetBstMidEndSepPunct{\mcitedefaultmidpunct}
{\mcitedefaultendpunct}{\mcitedefaultseppunct}\relax
\EndOfBibitem
\bibitem[Parzinger \latin{et~al.}(2017)Parzinger, Hetzl, Wurstbauer, and
  Holleitner]{parzinger_contact_2017}
Parzinger,~E.; Hetzl,~M.; Wurstbauer,~U.; Holleitner,~A.~W. \emph{npj {2D}
  Materials and Applications} \textbf{2017}, \emph{1}, 40\relax
\mciteBstWouldAddEndPuncttrue
\mciteSetBstMidEndSepPunct{\mcitedefaultmidpunct}
{\mcitedefaultendpunct}{\mcitedefaultseppunct}\relax
\EndOfBibitem
\bibitem[Rösner \latin{et~al.}(2016)Rösner, Steinke, Lorke, Gies, Jahnke, and
  Wehling]{rosner_two-dimensional_2016}
Rösner,~M.; Steinke,~C.; Lorke,~M.; Gies,~C.; Jahnke,~F.; Wehling,~T.~O.
  \emph{Nano Letters} \textbf{2016}, \emph{16}, 2322--2327\relax
\mciteBstWouldAddEndPuncttrue
\mciteSetBstMidEndSepPunct{\mcitedefaultmidpunct}
{\mcitedefaultendpunct}{\mcitedefaultseppunct}\relax
\EndOfBibitem
\bibitem[Steinke \latin{et~al.}(2017)Steinke, Mourad, Rösner, Lorke, Gies,
  Jahnke, Czycholl, and Wehling]{steinke_non-invasive_2017}
Steinke,~C.; Mourad,~D.; Rösner,~M.; Lorke,~M.; Gies,~C.; Jahnke,~F.;
  Czycholl,~G.; Wehling,~T.~O. \emph{Physical Review B} \textbf{2017},
  \emph{96}, {arXiv:} 1704.06095\relax
\mciteBstWouldAddEndPuncttrue
\mciteSetBstMidEndSepPunct{\mcitedefaultmidpunct}
{\mcitedefaultendpunct}{\mcitedefaultseppunct}\relax
\EndOfBibitem
\bibitem[Hong \latin{et~al.}(2017)Hong, Liu, Cao, Jin, Wang, Wang, and
  Liu]{hong_interfacial_2017}
Hong,~H.; Liu,~C.; Cao,~T.; Jin,~C.; Wang,~S.; Wang,~F.; Liu,~K. \emph{Advanced
  Materials Interfaces} \textbf{2017}, \emph{4}, n/a--n/a\relax
\mciteBstWouldAddEndPuncttrue
\mciteSetBstMidEndSepPunct{\mcitedefaultmidpunct}
{\mcitedefaultendpunct}{\mcitedefaultseppunct}\relax
\EndOfBibitem
\bibitem[Andersen \latin{et~al.}(2015)Andersen, Latini, and
  Thygesen]{andersen_dielectric_2015}
Andersen,~K.; Latini,~S.; Thygesen,~K.~S. \emph{Nano Letters} \textbf{2015},
  \emph{15}, 4616--4621\relax
\mciteBstWouldAddEndPuncttrue
\mciteSetBstMidEndSepPunct{\mcitedefaultmidpunct}
{\mcitedefaultendpunct}{\mcitedefaultseppunct}\relax
\EndOfBibitem
\bibitem[Rösner \latin{et~al.}(2015)Rösner, Şaşıoğlu, Friedrich, Blügel,
  and Wehling]{rosner_wannier_2015}
Rösner,~M.; Şaşıoğlu,~E.; Friedrich,~C.; Blügel,~S.; Wehling,~T.~O.
  \emph{Physical Review B} \textbf{2015}, \emph{92}, 085102\relax
\mciteBstWouldAddEndPuncttrue
\mciteSetBstMidEndSepPunct{\mcitedefaultmidpunct}
{\mcitedefaultendpunct}{\mcitedefaultseppunct}\relax
\EndOfBibitem
\bibitem[Latini \latin{et~al.}(2015)Latini, Olsen, and
  Thygesen]{latini_excitons_2015}
Latini,~S.; Olsen,~T.; Thygesen,~K.~S. \emph{Physical Review B} \textbf{2015},
  \emph{92}, 245123\relax
\mciteBstWouldAddEndPuncttrue
\mciteSetBstMidEndSepPunct{\mcitedefaultmidpunct}
{\mcitedefaultendpunct}{\mcitedefaultseppunct}\relax
\EndOfBibitem
\bibitem[Meckbach \latin{et~al.}(2017)Meckbach, Stroucken, and
  Koch]{meckbach_influence_2017}
Meckbach,~L.; Stroucken,~T.; Koch,~S.~W. \emph{{arXiv:1709.09056} [cond-mat]}
  \textbf{2017}, {arXiv:} 1709.09056\relax
\mciteBstWouldAddEndPuncttrue
\mciteSetBstMidEndSepPunct{\mcitedefaultmidpunct}
{\mcitedefaultendpunct}{\mcitedefaultseppunct}\relax
\EndOfBibitem
\bibitem[Steinhoff \latin{et~al.}(2014)Steinhoff, R\"osner, Jahnke, Wehling,
  and Gies]{steinhoff_influence_2014}
Steinhoff,~A.; R\"osner,~M.; Jahnke,~F.; Wehling,~T.~O.; Gies,~C. \emph{Nano
  Letters} \textbf{2014}, \emph{14}, 3743--3748\relax
\mciteBstWouldAddEndPuncttrue
\mciteSetBstMidEndSepPunct{\mcitedefaultmidpunct}
{\mcitedefaultendpunct}{\mcitedefaultseppunct}\relax
\EndOfBibitem
\bibitem[Sie \latin{et~al.}(2017)Sie, Steinhoff, Gies, Lui, Ma, Rösner,
  Schönhoff, Jahnke, Wehling, Lee, Kong, Jarillo-Herrero, and
  Gedik]{sie_observation_2017}
Sie,~E.~J.; Steinhoff,~A.; Gies,~C.; Lui,~C.~H.; Ma,~Q.; Rösner,~M.;
  Schönhoff,~G.; Jahnke,~F.; Wehling,~T.~O.; Lee,~Y.-H.; Kong,~J.;
  Jarillo-Herrero,~P.; Gedik,~N. \emph{Nano Letters} \textbf{2017}, \emph{17},
  4210--4216\relax
\mciteBstWouldAddEndPuncttrue
\mciteSetBstMidEndSepPunct{\mcitedefaultmidpunct}
{\mcitedefaultendpunct}{\mcitedefaultseppunct}\relax
\EndOfBibitem
\bibitem[Berghäuser and Malic(2014)Berghäuser, and
  Malic]{berghauser_analytical_2014}
Berghäuser,~G.; Malic,~E. \emph{Physical Review B} \textbf{2014}, \emph{89},
  125309\relax
\mciteBstWouldAddEndPuncttrue
\mciteSetBstMidEndSepPunct{\mcitedefaultmidpunct}
{\mcitedefaultendpunct}{\mcitedefaultseppunct}\relax
\EndOfBibitem
\bibitem[Rooney \latin{et~al.}(2017)Rooney, Kozikov, Rudenko, Prestat, Hamer,
  Withers, Cao, Novoselov, Katsnelson, Gorbachev, and
  Haigh]{rooney_observing_2017}
Rooney,~A.~P.; Kozikov,~A.; Rudenko,~A.~N.; Prestat,~E.; Hamer,~M.~J.;
  Withers,~F.; Cao,~Y.; Novoselov,~K.~S.; Katsnelson,~M.~I.; Gorbachev,~R.;
  Haigh,~S.~J. \emph{Nano Letters} \textbf{2017}, \relax
\mciteBstWouldAddEndPunctfalse
\mciteSetBstMidEndSepPunct{\mcitedefaultmidpunct}
{}{\mcitedefaultseppunct}\relax
\EndOfBibitem
\bibitem[Tongay \latin{et~al.}(2014)Tongay, Fan, Kang, Park, Koldemir, Suh,
  Narang, Liu, Ji, Li, Sinclair, and Wu]{tongay_tuning_2014}
Tongay,~S.; Fan,~W.; Kang,~J.; Park,~J.; Koldemir,~U.; Suh,~J.; Narang,~D.~S.;
  Liu,~K.; Ji,~J.; Li,~J.; Sinclair,~R.; Wu,~J. \emph{Nano Letters}
  \textbf{2014}, \emph{14}, 3185--3190\relax
\mciteBstWouldAddEndPuncttrue
\mciteSetBstMidEndSepPunct{\mcitedefaultmidpunct}
{\mcitedefaultendpunct}{\mcitedefaultseppunct}\relax
\EndOfBibitem
\bibitem[Hüser \latin{et~al.}(2013)Hüser, Olsen, and
  Thygesen]{huser_quasiparticle_2013}
Hüser,~F.; Olsen,~T.; Thygesen,~K.~S. \emph{Physical Review B} \textbf{2013},
  \emph{87}, 235132\relax
\mciteBstWouldAddEndPuncttrue
\mciteSetBstMidEndSepPunct{\mcitedefaultmidpunct}
{\mcitedefaultendpunct}{\mcitedefaultseppunct}\relax
\EndOfBibitem
\bibitem[Liu \latin{et~al.}(2013)Liu, Shan, Yao, Yao, and
  Xiao]{liu_three-band_2013}
Liu,~G.-B.; Shan,~W.-Y.; Yao,~Y.; Yao,~W.; Xiao,~D. \emph{Physical Review B}
  \textbf{2013}, \emph{88}, 085433\relax
\mciteBstWouldAddEndPuncttrue
\mciteSetBstMidEndSepPunct{\mcitedefaultmidpunct}
{\mcitedefaultendpunct}{\mcitedefaultseppunct}\relax
\EndOfBibitem
\bibitem[Tong \latin{et~al.}(2016)Tong, Yu, Zhu, Wang, Xu, and
  Yao]{tong_topological_2016}
Tong,~Q.; Yu,~H.; Zhu,~Q.; Wang,~Y.; Xu,~X.; Yao,~W. \emph{Nature Physics}
  \textbf{2016}, \emph{13}, nphys3968\relax
\mciteBstWouldAddEndPuncttrue
\mciteSetBstMidEndSepPunct{\mcitedefaultmidpunct}
{\mcitedefaultendpunct}{\mcitedefaultseppunct}\relax
\EndOfBibitem
\bibitem[Keldysh(1979)]{keldysh_coulomb_1979}
Keldysh,~L.~V. \emph{Soviet Journal of Experimental and Theoretical Physics
  Letters} \textbf{1979}, \emph{29}, 658\relax
\mciteBstWouldAddEndPuncttrue
\mciteSetBstMidEndSepPunct{\mcitedefaultmidpunct}
{\mcitedefaultendpunct}{\mcitedefaultseppunct}\relax
\EndOfBibitem
\bibitem[Haug and Koch(1993)Haug, and Koch]{haug_quantum_1993}
Haug,~H.; Koch,~S.~W. \emph{Quantum theory of the optical and electronic
  properties of semiconductors}; World Scientific: Singapore; River Edge, {NJ},
  1993\relax
\mciteBstWouldAddEndPuncttrue
\mciteSetBstMidEndSepPunct{\mcitedefaultmidpunct}
{\mcitedefaultendpunct}{\mcitedefaultseppunct}\relax
\EndOfBibitem
\bibitem[Rohlfing and Louie(2000)Rohlfing, and
  Louie]{rohlfing_electron-hole_2000}
Rohlfing,~M.; Louie,~S.~G. \emph{Physical Review B} \textbf{2000}, \emph{62},
  4927--4944\relax
\mciteBstWouldAddEndPuncttrue
\mciteSetBstMidEndSepPunct{\mcitedefaultmidpunct}
{\mcitedefaultendpunct}{\mcitedefaultseppunct}\relax
\EndOfBibitem
\bibitem[Steinhoff \latin{et~al.}(2017)Steinhoff, Florian, Rösner, Schönhoff,
  Wehling, and Jahnke]{steinhoff_exciton_2017}
Steinhoff,~A.; Florian,~M.; Rösner,~M.; Schönhoff,~G.; Wehling,~T.~O.;
  Jahnke,~F. \emph{Nature Communications} \textbf{2017}, \emph{8}, 1166\relax
\mciteBstWouldAddEndPuncttrue
\mciteSetBstMidEndSepPunct{\mcitedefaultmidpunct}
{\mcitedefaultendpunct}{\mcitedefaultseppunct}\relax
\EndOfBibitem
\bibitem[Olsen \latin{et~al.}(2016)Olsen, Latini, Rasmussen, and
  Thygesen]{olsen_simple_2016}
Olsen,~T.; Latini,~S.; Rasmussen,~F.; Thygesen,~K.~S. \emph{Physical Review
  Letters} \textbf{2016}, \emph{116}, 056401\relax
\mciteBstWouldAddEndPuncttrue
\mciteSetBstMidEndSepPunct{\mcitedefaultmidpunct}
{\mcitedefaultendpunct}{\mcitedefaultseppunct}\relax
\EndOfBibitem
\bibitem[Rohlfing(2010)]{rohlfing_electronic_2010}
Rohlfing,~M. \emph{Physical Review B} \textbf{2010}, \emph{82}, 205127\relax
\mciteBstWouldAddEndPuncttrue
\mciteSetBstMidEndSepPunct{\mcitedefaultmidpunct}
{\mcitedefaultendpunct}{\mcitedefaultseppunct}\relax
\EndOfBibitem
\bibitem[Winther and Thygesen(2017)Winther, and Thygesen]{winther_band_2017}
Winther,~K.~T.; Thygesen,~K.~S. \emph{{2D} Materials} \textbf{2017}, \emph{4},
  025059\relax
\mciteBstWouldAddEndPuncttrue
\mciteSetBstMidEndSepPunct{\mcitedefaultmidpunct}
{\mcitedefaultendpunct}{\mcitedefaultseppunct}\relax
\EndOfBibitem
\bibitem[Mak \latin{et~al.}(2013)Mak, He, Lee, Lee, Hone, Heinz, and
  Shan]{mak_tightly_2013}
Mak,~K.~F.; He,~K.; Lee,~C.; Lee,~G.~H.; Hone,~J.; Heinz,~T.~F.; Shan,~J.
  \emph{Nature Materials} \textbf{2013}, \emph{12}, 207--211\relax
\mciteBstWouldAddEndPuncttrue
\mciteSetBstMidEndSepPunct{\mcitedefaultmidpunct}
{\mcitedefaultendpunct}{\mcitedefaultseppunct}\relax
\EndOfBibitem
\bibitem[Ross \latin{et~al.}(2013)Ross, Wu, Yu, Ghimire, Jones, Aivazian, Yan,
  Mandrus, Xiao, Yao, and Xu]{ross_electrical_2013}
Ross,~J.~S.; Wu,~S.; Yu,~H.; Ghimire,~N.~J.; Jones,~A.~M.; Aivazian,~G.;
  Yan,~J.; Mandrus,~D.~G.; Xiao,~D.; Yao,~W.; Xu,~X. \emph{Nature
  Communications} \textbf{2013}, \emph{4}, 1474\relax
\mciteBstWouldAddEndPuncttrue
\mciteSetBstMidEndSepPunct{\mcitedefaultmidpunct}
{\mcitedefaultendpunct}{\mcitedefaultseppunct}\relax
\EndOfBibitem
\bibitem[Scheuschner \latin{et~al.}(2014)Scheuschner, Ochedowski, Kaulitz,
  Gillen, Schleberger, and Maultzsch]{scheuschner_photoluminescence_2014}
Scheuschner,~N.; Ochedowski,~O.; Kaulitz,~A.-M.; Gillen,~R.; Schleberger,~M.;
  Maultzsch,~J. \emph{Physical Review B} \textbf{2014}, \emph{89}, 125406\relax
\mciteBstWouldAddEndPuncttrue
\mciteSetBstMidEndSepPunct{\mcitedefaultmidpunct}
{\mcitedefaultendpunct}{\mcitedefaultseppunct}\relax
\EndOfBibitem
\bibitem[Ulstrup \latin{et~al.}(2016)Ulstrup, Čabo, Miwa, Riley, Grønborg,
  Johannsen, Cacho, Alexander, Chapman, Springate, Bianchi, Dendzik, Lauritsen,
  King, and Hofmann]{ulstrup_ultrafast_2016}
Ulstrup,~S.; Čabo,~A.~G.; Miwa,~J.~A.; Riley,~J.~M.; Grønborg,~S.~S.;
  Johannsen,~J.~C.; Cacho,~C.; Alexander,~O.; Chapman,~R.~T.; Springate,~E.;
  Bianchi,~M.; Dendzik,~M.; Lauritsen,~J.~V.; King,~P. D.~C.; Hofmann,~P.
  \emph{{ACS} Nano} \textbf{2016}, \emph{10}, 6315--6322\relax
\mciteBstWouldAddEndPuncttrue
\mciteSetBstMidEndSepPunct{\mcitedefaultmidpunct}
{\mcitedefaultendpunct}{\mcitedefaultseppunct}\relax
\EndOfBibitem
\bibitem[Esser \latin{et~al.}(2001)Esser, Zimmermann, and
  Runge]{esser_theory_2001}
Esser,~A.; Zimmermann,~R.; Runge,~E. \emph{physica status solidi (b)}
  \textbf{2001}, \emph{227}, 317--330\relax
\mciteBstWouldAddEndPuncttrue
\mciteSetBstMidEndSepPunct{\mcitedefaultmidpunct}
{\mcitedefaultendpunct}{\mcitedefaultseppunct}\relax
\EndOfBibitem
\bibitem[Berkelbach \latin{et~al.}(2013)Berkelbach, Hybertsen, and
  Reichman]{berkelbach_theory_2013}
Berkelbach,~T.~C.; Hybertsen,~M.~S.; Reichman,~D.~R. \emph{Physical Review B}
  \textbf{2013}, \emph{88}, 045318\relax
\mciteBstWouldAddEndPuncttrue
\mciteSetBstMidEndSepPunct{\mcitedefaultmidpunct}
{\mcitedefaultendpunct}{\mcitedefaultseppunct}\relax
\EndOfBibitem
\bibitem[Mayers \latin{et~al.}(2015)Mayers, Berkelbach, Hybertsen, and
  Reichman]{mayers_binding_2015}
Mayers,~M.~Z.; Berkelbach,~T.~C.; Hybertsen,~M.~S.; Reichman,~D.~R.
  \emph{Physical Review B} \textbf{2015}, \emph{92}, 161404\relax
\mciteBstWouldAddEndPuncttrue
\mciteSetBstMidEndSepPunct{\mcitedefaultmidpunct}
{\mcitedefaultendpunct}{\mcitedefaultseppunct}\relax
\EndOfBibitem
\bibitem[Kylänpää and Komsa(2015)Kylänpää, and
  Komsa]{kylanpaa_binding_2015}
Kylänpää,~I.; Komsa,~H.-P. \emph{Physical Review B} \textbf{2015},
  \emph{92}, 205418\relax
\mciteBstWouldAddEndPuncttrue
\mciteSetBstMidEndSepPunct{\mcitedefaultmidpunct}
{\mcitedefaultendpunct}{\mcitedefaultseppunct}\relax
\EndOfBibitem
\bibitem[Velizhanin and Saxena(2015)Velizhanin, and
  Saxena]{velizhanin_excitonic_2015}
Velizhanin,~K.~A.; Saxena,~A. \emph{Physical Review B} \textbf{2015},
  \emph{92}, 195305\relax
\mciteBstWouldAddEndPuncttrue
\mciteSetBstMidEndSepPunct{\mcitedefaultmidpunct}
{\mcitedefaultendpunct}{\mcitedefaultseppunct}\relax
\EndOfBibitem
\bibitem[Ganchev \latin{et~al.}(2015)Ganchev, Drummond, Aleiner, and
  Fal’ko]{ganchev_three-particle_2015}
Ganchev,~B.; Drummond,~N.; Aleiner,~I.; Fal’ko,~V. \emph{Physical Review
  Letters} \textbf{2015}, \emph{114}, 107401\relax
\mciteBstWouldAddEndPuncttrue
\mciteSetBstMidEndSepPunct{\mcitedefaultmidpunct}
{\mcitedefaultendpunct}{\mcitedefaultseppunct}\relax
\EndOfBibitem
\bibitem[Kidd \latin{et~al.}(2016)Kidd, Zhang, and Varga]{kidd_binding_2016}
Kidd,~D.~W.; Zhang,~D.~K.; Varga,~K. \emph{Physical Review B} \textbf{2016},
  \emph{93}, 125423\relax
\mciteBstWouldAddEndPuncttrue
\mciteSetBstMidEndSepPunct{\mcitedefaultmidpunct}
{\mcitedefaultendpunct}{\mcitedefaultseppunct}\relax
\EndOfBibitem
\bibitem[Szyniszewski \latin{et~al.}(2017)Szyniszewski, Mostaani, Drummond, and
  Fal'ko]{szyniszewski_binding_2017}
Szyniszewski,~M.; Mostaani,~E.; Drummond,~N.~D.; Fal'ko,~V.~I. \emph{Physical
  Review B} \textbf{2017}, \emph{95}, 081301\relax
\mciteBstWouldAddEndPuncttrue
\mciteSetBstMidEndSepPunct{\mcitedefaultmidpunct}
{\mcitedefaultendpunct}{\mcitedefaultseppunct}\relax
\EndOfBibitem
\bibitem[Wierzbowski \latin{et~al.}(2017)Wierzbowski, Klein, Sigger,
  Straubinger, Kremser, Taniguchi, Watanabe, Wurstbauer, Holleitner, Kaniber,
  Müller, and Finley]{wierzbowski_direct_2017}
Wierzbowski,~J.; Klein,~J.; Sigger,~F.; Straubinger,~C.; Kremser,~M.;
  Taniguchi,~T.; Watanabe,~K.; Wurstbauer,~U.; Holleitner,~A.~W.; Kaniber,~M.;
  Müller,~K.; Finley,~J.~J. \emph{Scientific Reports} \textbf{2017}, \emph{7},
  12383\relax
\mciteBstWouldAddEndPuncttrue
\mciteSetBstMidEndSepPunct{\mcitedefaultmidpunct}
{\mcitedefaultendpunct}{\mcitedefaultseppunct}\relax
\EndOfBibitem
\bibitem[Ajayi \latin{et~al.}(2017)Ajayi, Ardelean, Shepard, Wang, Antony,
  {{Takeshi} Taniguchi}, Watanabe, Heinz, Strauf, Zhu, and
  Hone]{ajayi_approaching_2017}
Ajayi,~O.~A.; Ardelean,~J.~V.; Shepard,~G.~D.; Wang,~J.; Antony,~A.; {{Takeshi}
  Taniguchi},; Watanabe,~K.; Heinz,~T.~F.; Strauf,~S.; Zhu,~X.-Y.; Hone,~J.~C.
  \emph{{2D} Materials} \textbf{2017}, \emph{4}, 031011\relax
\mciteBstWouldAddEndPuncttrue
\mciteSetBstMidEndSepPunct{\mcitedefaultmidpunct}
{\mcitedefaultendpunct}{\mcitedefaultseppunct}\relax
\EndOfBibitem
\bibitem[Cadiz \latin{et~al.}(2017)Cadiz, Courtade, Robert, Wang, Shen, Cai,
  Taniguchi, Watanabe, Carrere, Lagarde, Manca, Amand, Renucci, Tongay, Marie,
  and Urbaszek]{cadiz_excitonic_2017}
Cadiz,~F. \latin{et~al.}  \emph{Physical Review X} \textbf{2017}, \emph{7},
  021026\relax
\mciteBstWouldAddEndPuncttrue
\mciteSetBstMidEndSepPunct{\mcitedefaultmidpunct}
{\mcitedefaultendpunct}{\mcitedefaultseppunct}\relax
\EndOfBibitem
\bibitem[Wang \latin{et~al.}(2016)Wang, Huang, Tian, Ceballos, Lin,
  Mahjouri-Samani, Boulesbaa, Puretzky, Rouleau, Yoon, Zhao, Xiao, Duscher, and
  Geohegan]{wang_interlayer_2016}
Wang,~K.; Huang,~B.; Tian,~M.; Ceballos,~F.; Lin,~M.-W.; Mahjouri-Samani,~M.;
  Boulesbaa,~A.; Puretzky,~A.~A.; Rouleau,~C.~M.; Yoon,~M.; Zhao,~H.; Xiao,~K.;
  Duscher,~G.; Geohegan,~D.~B. \emph{{ACS} Nano} \textbf{2016}, \emph{10},
  6612--6622\relax
\mciteBstWouldAddEndPuncttrue
\mciteSetBstMidEndSepPunct{\mcitedefaultmidpunct}
{\mcitedefaultendpunct}{\mcitedefaultseppunct}\relax
\EndOfBibitem
\bibitem[Borghardt \latin{et~al.}(2017)Borghardt, Tu, Winkler, Schubert,
  Zander, Leosson, and Kardynał]{borghardt_engineering_2017}
Borghardt,~S.; Tu,~J.-S.; Winkler,~F.; Schubert,~J.; Zander,~W.; Leosson,~K.;
  Kardynał,~B.~E. \emph{Physical Review Materials} \textbf{2017}, \emph{1},
  054001\relax
\mciteBstWouldAddEndPuncttrue
\mciteSetBstMidEndSepPunct{\mcitedefaultmidpunct}
{\mcitedefaultendpunct}{\mcitedefaultseppunct}\relax
\EndOfBibitem
\bibitem[Lee \latin{et~al.}(2016)Lee, Kim, Kim, and Lee]{lee_identifying_2016}
Lee,~H.~S.; Kim,~M.~S.; Kim,~H.; Lee,~Y.~H. \emph{Physical Review B}
  \textbf{2016}, \emph{93}, 140409\relax
\mciteBstWouldAddEndPuncttrue
\mciteSetBstMidEndSepPunct{\mcitedefaultmidpunct}
{\mcitedefaultendpunct}{\mcitedefaultseppunct}\relax
\EndOfBibitem
\end{mcitethebibliography}
\end{document}